\documentclass[12pt]{article}

\usepackage{amsmath}
\usepackage{graphicx}
\usepackage{amsfonts}
\usepackage{amssymb}
\usepackage{epsfig}
\usepackage{color}
\usepackage{psfrag}

\newcommand{\EQ}{\begin{equation}}
\newcommand{\EN}{\end{equation}}

\begin{document}
\setcounter{page}{0} \topmargin 0pt \oddsidemargin 5mm
\renewcommand{\thefootnote}{\arabic{footnote}}
\newpage
\setcounter{page}{0}

\begin{titlepage}

\begin{flushright}
\end{flushright}
\vspace{0.5cm}
\begin{center}
{\Large {\bf Kinks and Particles in Non-integrable}}\\
{\Large {\bf Quantum Field Theories}}\\
\vspace{2cm}
{\large G. Mussardo$^{a,b}$}
\\
\vspace{0.5cm} $^a${\em International School for Advanced Studies} \\
{\em  Via Beirut 1, 34013 Trieste, Italy}\\
\vspace{0.3cm} $^b${\em Istituto Nazionale di Fisica Nucleare, 
Sezione di Trieste}

\end{center}
\vspace{1cm}

\begin{abstract}
\noindent
In this talk we discuss an elementary derivation of the 
semi-classical spectrum 
of neutral particles in two field theories with kink excitations. We also show that, 
in the non-integrable cases, each vacuum state cannot generically support more than two stable 
particles, since all other neutral exitations are resonances, which will eventually decay.

\end{abstract}

\end{titlepage}

\newpage

\section{Introduction}\label{intro}

Two--dimensional massive Integrable Quantum Field Theories (IQFTs)  
have proven to be one of the most successful topics of relativistic 
field theory, 
with a large variety of applications to statistical mechanical models. 
The main reason for this success consists of their simplified on--shell  
dynamics which is encoded into a set of elastic and factorized scattering 
amplitudes of their massive particles \cite{zamzam,Zam}. 
The two-particle S-matrix has a very simple analytic structure, with only 
poles in the physical strip, and it can be computed combining 
the standard requirements of unitarity, crossing and factorization 
together with specific symmetry properties of the theory.
The complete mass spectrum is obtained looking at the pole singularities 
of the $S$--matrix elements. Off--mass shell quantities, such as the 
correlation functions, can be also determined once the elastic $S$--matrix 
and the mass spectrum are known. In fact, one can compute the exact matrix 
elements of the (semi)local fields on the asymptotic states with the 
Form Factor (FF) approach \cite{Smirnov}, and use them to write down the spectral 
representation of the correlators. By following this approach, it has been 
possible, for instance, to tackle successfully the long-standing problem 
of spelling out the mass spectrum and the correlation functions of the 
two dimensional Ising model in a magnetic field
\cite{Zam,DMIsing}, as well as many other interesting problems 
of statistical physics (for a partial list of them see, for instance, 
\cite{GM}). 

The S-matrix approach can be also constructed for massless IQFTs 
\cite{zam.massless,zam-zam.massless,fsw,dms2}, despite the subtleties 
in defining a scattering theory between massless particles in $(1+1)$ 
dimensions, and turns out to be useful mainly when conformal symmetry 
is not present. In this case, massless  IQFTs generically describe the 
Renormalization Group trajectories connecting two different Conformal 
Field Theories, which respectively rule the ultraviolet 
and infrared limits of all physical quantities along the flows. 

Given the large number of remarkable results obtained by the study of 
IQFTs, one of the most interesting challenges is to extend 
the analysis to the non--integrable field theories, at least to those 
obtained as deformations of the integrable ones and to develop the 
corresponding perturbation theory. The breaking of integrability 
is expected to considerably increase the difficulties of the mathematical 
analysis, since scattering processes are no longer elastic. Non--integrable 
field theories are in fact generally characterized by particle production 
amplitudes, resonance states and, correspondingly, decay events.
All these features strongly effect the analytic structure of the 
scattering amplitudes, introducing a rich pattern of branch cut singularities,
in addition to the pole structure associated to bound and resonance states. 
For massive non--integrable field theories, a convenient perturbative scheme 
was originally proposed in \cite{dms} and called Form Factor Perturbation Theory 
(FFPT), since it is based on the knowledge of the exact Form Factors (FFs) 
of the original integrable theory. It was shown that, even using just the first 
order correction of the FFPT, a great deal of information can be obtained,  
such as the evolution of their particle content, the variation of their masses 
and the change of the ground state energy. Whenever possible, universal ratios 
were computed and successfully compared with their value obtained by 
other means. Recently, for instance, it has been obtained the universal ratios 
relative to the decay of the particles with higher masses in the Ising model in 
a magnetic field, once the temperature is displayed away from the critical value \cite{grinza} 
(see also the contibution by G. Delfino in this proceedings \cite{Aldo}). 
For other and important aspects of the Ising model along non-integrable lines see the 
references \cite{mccoy,fonsecazam,rut,zamfon}.  
Applied to the double Sine--Gordon model \cite{dm}, the FFPT has been useful 
in clarifying the rich dynamics of this non--integrable model. 
In particular, in  relating the confinement of the kinks in the deformed theory 
to the non--locality properties of the perturbed operator and predicting the 
existence of a Ising--like phase transition for particular ratios of the two 
frequencies -- results which were later confirmed by a numerical 
study \cite{takacs}. The FFPT has been also used to study the spectrum of the 
$O(3)$ non-linear sigma model with a topological $\theta$ term, by varying $\theta$ 
\cite{CMPRL,conmus}. 

In this talk I would like to focus the attention on a different approach to 
tackle some interesting non-integrable models, i.e. those two dimensional field theories 
with kink topological excitations. Such theories are
described by a scalar real field $\varphi(x)$, with a Lagrangian density 
\EQ
{\cal L} \,=\,\frac{1}{2} (\partial_{\mu} \varphi)^2 - U(\varphi) \,\,\,, 
\label{Lagrangian}
\EN 
where the potential $U(\varphi)$ possesses several degenerate minima at 
$\varphi_a^{(0)}$ ($a =1,2,\ldots,n$), as the one shown in Figure 1. These 
minima correspond to the different vacua $\mid \,a\,\rangle$ of 
the associate quantum field theory. 

\vspace{5mm}


\begin{figure}[h]
\begin{tabular}{p{8cm}p{8cm}}
\psfig{figure=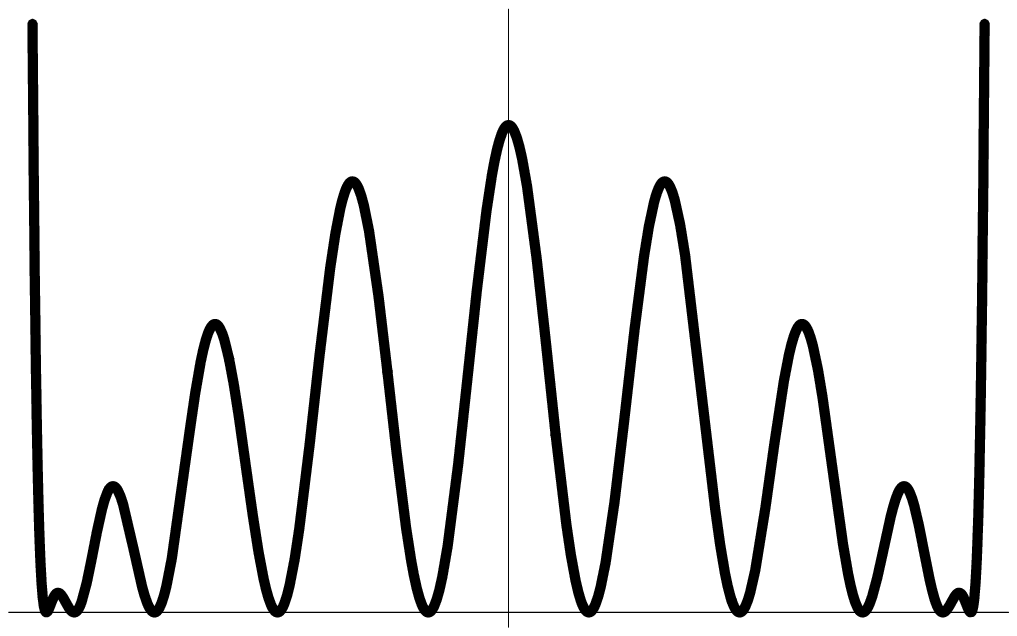,
height=4cm,width=6cm} \vspace{0.2cm}&
\psfig{figure=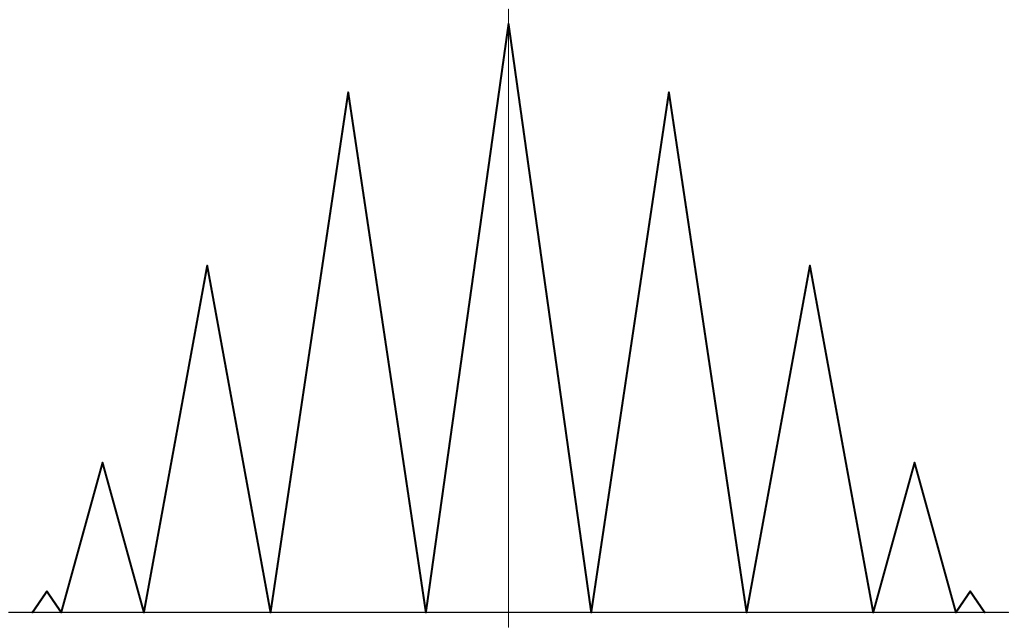,height=4cm,width=6cm} \\
\hspace{2.7cm}(A)  & \hspace{2.7cm} (B) 
\end{tabular}
\caption{{\em Potential $U(\varphi)$ of a quantum field theory with kink 
excitations (A) and istogram of the masses of the kinks (B).}} 
\label{potential}
\end{figure}

\noindent
The basic excitations of this kind of models are kinks and anti-kinks, i.e. topological 
configurations which interpolate between two neighbouring vacua. Semiclassically they 
correspond to the static solutions of the equation of motion, i.e. 
\EQ
\partial^2_x \,\varphi(x) \,=\,U'[\varphi(x)] \,\,\,, 
\label{static}
\EN
with boundary conditions $\varphi(-\infty) = \varphi_a^{(0)}$ and 
$\varphi(+\infty)= \varphi_{b}^{(0)}$, where $b = a \pm 1$. 
Denoting by $\varphi_{ab}(x)$ the solutions of this equation, their 
classical energy density is given by 
\EQ
\epsilon_{ab}(x) \,=\,\frac{1}{2} \left(\frac{d\varphi_{ab}}{d x}\right)^2 + U(\varphi_{ab}(x))  \,\,\,,
\EN 
and its integral provides the classical expression of the kink masses 
\EQ
M_{ab} \,=\,\int_{-\infty}^{\infty} \epsilon_{ab}(x) \,\,\,.
\label{integralmass}
\EN   
It is easy to show that the classical masses of the kinks $\varphi_{ab}(x)$ are 
simply proportional to the heights of the potential between the two minima 
$\varphi_a^{(0)}$ and $\varphi_b^{(0)}$: their istogram provides a caricature 
of the original ptential (see Figura 1).  

\begin{figure}[h]
\hspace{10mm}
\vspace{8mm}
\psfig{figure=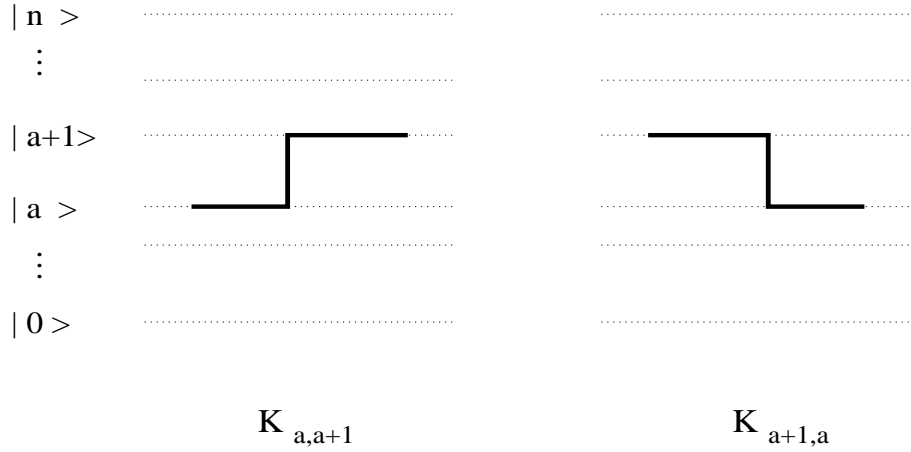,height=6cm,width=12cm}
\caption{{\em Kink and antikink configurations.}}
\label{step}
\end{figure}

The classical solutions can be set in motion by a Lorentz transformation, i.e. 
$\varphi_{ab}(x) \rightarrow \varphi_{ab}\left[(x \pm v t)/\sqrt{1-v^2}\right]$. 
In the quantum theory, these configurations describe the kink states $\mid K_{ab}(\theta)\,\rangle$, 
where $a$ and $b$ are the indices of the initial and final vacuum, respectively. 
The quantity $\theta$ is the rapidity variable which parameterises the relativistic 
dispersion relation of these excitations, i.e. 
\EQ
E = M_{ab}\,\cosh\theta
\,\,\,\,\,\,\,
,
\,\,\,\,\,\,\,
P = M_{ab} \,\sinh\theta
\,\,\,.
\label{rapidity}
\EN 
Conventionally $\mid K_{a,a+1}(\theta) \,\rangle$ denotes the {\it kink} between the pair 
of vacua $\left\{\mid a\,\rangle ,\mid a+1\,\rangle\right\}$ while $\mid K_{a+1,a}\,\rangle$ 
is the corresponding {\it anti-kink}. For the kink configurations it may be useful 
to adopt the simplified graphical form shown in Figure \ref{step}.

\vspace{3mm}

\noindent
The multi-particle states are given by a string of these excitations, with the adjacency 
condition of the consecutive indices for the continuity of the field configuration
\EQ
\mid K_{a_1,a_2}(\theta_1) \,K_{a_2,a_3}(\theta_2)\,K_{a_3,a_4}(\theta_3) \ldots \rangle 
\,\,\,\,\,\,\,\,
,
\,\,\,\,\,\,\,\, (a_{i+1} = a_i \pm 1)
\EN 
In addition to the kinks, in the quantum theory there may exist other excitations in the guise of 
ordinary scalar particles (breathers). These are the neutral excitations $\mid B_c(\theta)\,\rangle_a$ 
($c=1,2,\ldots$) around each of the vacua $\mid a\,\rangle$. For a theory based on a Lagrangian of 
a single real field, these states are all non-degenerate: in fact, there are no extra quantities 
which commute with the Hamiltonian and that can give rise to a multiplicity of them. The only exact 
(alias, unbroken) symmetries for a Lagrangian as (\ref{Lagrangian}) may be the discrete ones, like the 
parity transformation $P$, for instance, or the charge conjugation ${\cal C}$. However, since they 
are neutral excitations, they will be either even or odd eigenvectors of ${\cal C}$. 
 
\begin{figure}[h]
\hspace{10mm}
\vspace{8mm}
\psfig{figure=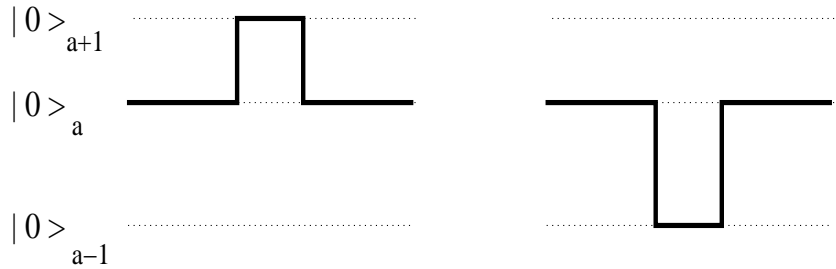,height=35mm,width=11cm}
\caption{{\em Kink-antikink configurations which may give rise to a bound state nearby the vacuum 
$\mid 0\,\rangle_a$.}}
\label{tooth}
\end{figure}

The neutral particles must be identified as the bound states of the kink-antikink configurations that 
start and end at the same vacuum $\mid a\,\rangle$, i.e. $\mid K_{ab}(\theta_1) \,K_{ba}(\theta_2)\,
\rangle$, with the ``tooth'' shapes shown in Figure\,\,\ref{tooth}.

\vspace{3mm}

\noindent
If such two-kink states have a pole at an imaginary value 
$i \,u_{a b}^c$ within the physical strip $0 < {\rm Im}\, \theta < \pi$ of their rapidity difference 
$\theta = \theta_1 - \theta_2$, then their bound states are defined through the factorization formula 
which holds in the vicinity of this singularity 
\EQ
\mid K_{ab}(\theta_1) \,K_{ba}(\theta_2) \,\rangle \,\simeq \,i\,\frac{g_{ab}^c}{\theta - i u_{ab}^c}
\,\mid B_c\,\rangle_a \,\,\,.
\label{factorization}
\EN 
In this expression $g_{ab}^c$ is the on-shell 3-particle coupling between the kinks and the 
neutral particle. Moreover, the mass of the bound states is simply obtained by substituing 
the resonance value $i \,u_{ab}^c$ within the expression of the Mandelstam variable $s$ 
of the two-kink channel 
\EQ
s = 4 M^2_{ab} \,\cosh^2\frac{\theta}{2} 
\,\,\,\,\,\,
\longrightarrow 
\,\,\,\,\,\, 
m_c \,=\,2 M_{ab} \cos\frac{u_{ab}^c}{2} \,\,\,.
\label{massboundstate}
\EN 

Concerning the vacua themselves, as well known, in the infinite volume their classical 
degeneracy is removed by selecting one of them, say  $\mid k \,\rangle$, out of the $n$ 
available. This happens through the usual spontaneously symmetry breaking mechanism, even 
though -- stricly speaking -- there may be no internal symmetry to break at all. This is 
the case, for instance, of the potential shown in Figure 1, which does not have any 
particular invariance. In the absence of a symmetry which connects the various 
vacua, the world -- as seen by each of them -- may appear very different: they can have, 
indeed, different particle contents. The problem we would like to examine in this talk 
concerns the neutral excitations around each vacuum, in particular the question of the 
existence of such particles and of the value of their masses. To this aim, let's make 
use of a semiclassical approach.

\section{A semiclassical formula}

\begin{figure}[h]
\hspace{55mm}
\vspace{1mm}
\psfig{figure=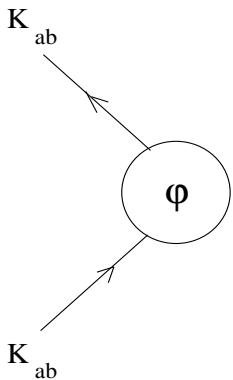,height=5cm,width=3cm}
\caption{{\em Matrix element between kink states.}}
\label{formfactor}
\end{figure}

The starting point of our analysis is a remarkably simple formula due to Goldstone-Jackiw \cite{GJ}, 
which is valid in the semiclassical approximation, i.e. when the coupling constant goes to zero and 
the mass of the kinks becomes correspondingly very large with respect to any other mass scale.  
In its refined version, given in \cite{JW} and rediscovered in \cite{FFvolume}, 
it reads as follows\footnote{The matrix element of the field $\varphi(y)$ is easily obtained by using 
$\varphi(y) = e^{-i P_{\mu} y^{\mu}} \,\varphi(0) \,e^{i P_{\mu} y^{\mu}}$ and by acting with the 
conserved energy-momentum operator $P_{\mu}$ on the kink state. Moreover, for the semiclassical 
matrix element $F_{ab}^{\cal G}(\theta)$ of the operator $G[\varphi(0)]$, one should employ 
$G[\varphi_{ab}(x)]$. For instance, the matrix element of $\varphi^2(0)$  
are given by the Fourier transform of $\varphi_{ab}^2(x)$.} (Figure \ref{formfactor}) 
\EQ
f_{ab}^{\varphi}(\theta) \,=\,\langle K_{ab}(\theta_1) \,\mid \varphi(0) \,\mid \, 
K_{ab}(\theta_2) \rangle 
\,\simeq \,\int_{-\infty}^{\infty} dx \,e^{i M_{ab} \,\theta\,x} \,\,
\varphi_{ab}(x) \,\,\,,
\label{remarkable1}
\EN  
where $\theta = \theta_1 - \theta_2$. 

\vspace{3mm}

\noindent
Notice that, if we substitute in the above formula 
$\theta \rightarrow i \pi - \theta$, the corresponding expression may be interpreted as 
the following Form Factor 
\EQ
F_{ab}^{\varphi}(\theta) \,=\, f(i \pi - \theta) \,=\,\langle a \,\mid \varphi(0) \,\mid \,
K_{ab}(\theta_1) \, K_{ba}(\theta_2) \rangle \,\,\,.
\label{remarkable2}
\EN    
In this matrix element, it appears the neutral kink states around the vacuum $\mid a \rangle$ 
we are interested in. 

Eq.\,(\ref{remarkable1}) deserves several comments. 

\begin{enumerate}
\item 
The appealing aspect of the formula (\ref{remarkable1}) stays in the relation between the 
Fourier transform of the {\em classical} configuration of the kink, -- i.e. the solution 
$\varphi_{ab}(x)$ of the differential equation (\ref{static}) -- to the {\em quantum} 
matrix element of the field $\varphi(0)$ between the vacuum $\mid a\,\rangle$ and the 
2-particle kink state $\mid K_{ab}(\theta_1) \,K_{ba}(\theta_2)\,\rangle$. 

Once the solution of eq.\,(\ref{static}) has been found and its Fourier transform has 
been taken, the poles of $F_{ab}(\theta)$ within the physical strip of $\theta$ identify the 
neutral bound states which couple to $\varphi$. The mass of the neutral 
particles can be extracted by using eq.\,(\ref{massboundstate}), while the on-shell 3-particle 
coupling $g_{ab}^c$ can be obtained from the residue at these poles (Figura \ref{residuef})  
\EQ
\lim_{\theta \rightarrow i \,u_{ab}^c} (\theta - i u_{ab}^c)\, F_{ab}(\theta)
\,=\,i \,g_{ab}^c \,\,\langle a \,\mid \varphi(0) \,\mid \,B_c \,\rangle \,\,\,.
\label{residue}
\EN  

\begin{figure}[h]
\hspace{15mm}
\vspace{10mm}
\psfig{figure=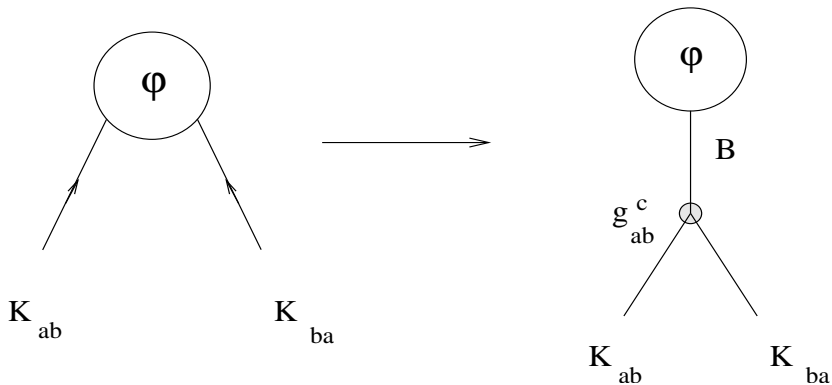,height=5cm,width=11cm}
\vspace{1mm}
\caption{{\em Residue equation for the matrix element on the kink states.}}
\label{residuef}
\end{figure}

\item It is important to stress that, for a generic theory, the classical kink configuration $\varphi_{ab}(x)$ is 
not related in a simple way to the anti-kink configuration $\varphi_{ba}(x)$. It is precisely for this reason 
that neighbouring vacua may have a different spectrum of neutral excitations, as shown in the examples 
discussed in the following sections.   

\item It is also worth noting that this procedure for extracting the bound states masses permits in many cases 
to avoid the semiclassical quantization of the breather solutions \cite{DHN}, making their derivation much 
simpler. The reason is that, the classical breather configurations depend also on time and have, in general, 
a more complicated structure than the kink ones. Yet, it can be shown that in non--integrable theories 
these configurations do not exist as exact solutions of the partial differential equations of the field 
theory. On the contrary, in order to apply eq.\,(\ref{remarkable1}), one simply needs the solution of 
the {\em ordinary} differential equation (\ref{static}). It is worth notice that, to locate the 
poles of $f_{ab}^{\varphi}(\theta)$, one only needs to looking at the exponential behavior of the classical 
solutions at $x \rightarrow \pm \infty$, as discussed below.

\end{enumerate}

In the next two sections we will present the analyse a class 
of theories with only two vacua, which can be either symmetric or asymmetric ones. A complete 
analysis of other potentials can be found in the original paper \cite{kink}.  

\section{Symmetric wells}\label{phi4section}

A prototype example of a potential with two symmetric wells is the $\varphi^4$ theory in its 
broken phase. The potential is given in this case by 
\EQ
U(\varphi) \,=\,\frac{\lambda}{4} \left(\varphi^2 - \frac{m^2}{\lambda}\right)^2 \,\,\,.
\EN 
Let us denote with $\mid \pm 1 \,\rangle$ the vacua corresponding to the classical 
minima $\varphi_{\pm}^{(0)} \,=\,\pm \frac{m}{\sqrt{\lambda}}$. By expanding around them, 
$\varphi \,=\,\varphi_{\pm}^{(0)} + \eta$, we have 
\EQ
U(\varphi_{\pm}^{(0)} + \eta) \,=\, m^2 \,\eta^2 \pm m \sqrt{\lambda}\,\eta^3 + \frac{\lambda}{4}
\eta^4 \,\,\,.
\label{potentialphi4}
\EN 
Hence, perturbation theory predicts the existence of a neutral particle for each of the two vacua,  
with a bare mass given by $m_b = \sqrt{2} m$, irrespectively of the value of the coupling $\lambda$. 
Let's see, instead, what is the result of the semiclassical analysis. 

The kink solutions are given in this case by
\EQ
\varphi_{-a,a}(x) \,=\,a \,\frac{m}{\sqrt{\lambda}} \,\tanh\left[\frac{m x}{\sqrt{2}}\right]
\,\,\,\,\,\,\,
,
\,\,\,\,\,\,\, a = \pm 1 
\label{kinksolphi4}
\EN 
and their classical mass is 
\EQ
M_0\,=\,\int_{-\infty}^{\infty} \epsilon(x) \,dx \,=\,\frac{2 \sqrt{2}}{3} \,
\frac{m^3}{\lambda}  
\,\,\,.
\EN 
The value of the potential at the origin, which gives the height of the barrier between the two vacua, 
can be expressed as 
\EQ
U(0) \,=\,\frac{3 m}{8 \sqrt{2}} \,M_0 \,\,\,,
\EN 
and, as noticed in the introduction, is proportional to the classical mass of the kink. 

If we take into account the contribution of the small oscillations around the classical static 
configurations, the kink mass gets corrected as \cite{DHN}
\EQ
M \,=\,\frac{2 \sqrt{2}}{3} \,\frac{m^3}{\lambda} - m 
\left(\frac{3}{\pi \sqrt{2}} - \frac{1}{2 \sqrt{6}}\right) 
+ {\cal O}(\lambda) \,\,\,.
\label{mass1phi4}
\EN 
It is convenient to define 
\[
c =  \left(\frac{3}{2\pi} - \frac{1}{4 \sqrt{3}}\right) > 0 \,\,\,,
\]
and also the adimensional quantities
\EQ
g = \frac{3 \lambda}{2 \pi m^2}
\,\,\,\,\,\,\,\,
;
\,\,\,\,\,\,\,\,
\xi \,=\,\frac{g}{1 - \pi c g} \,\,\,.
\label{definitiong}
\EN 
In terms of them, the mass of the kink can be expressed as 
\EQ
M \,=\,\frac{\sqrt{2} m}{\pi \,\xi}\,=\,\frac{m_b}{\pi \,\xi}\,\,\,.
\label{newmassphi4}
\EN
Since the kink and the anti-kink solutions are equal functions (up to a sign), their Fourier transforms 
have the same poles. Hence, the spectrum of the neutral particles will be the same on both vacua, in 
agreement with the $Z_2$ symmetry of the model. We have  
\EQ
f_{-a,a}(\theta) \,= \, \int_{-\infty}^{\infty} 
dx \,e^{i M  \theta\,x} \varphi_{-a,a}(x)  \nonumber \, =\, 
 i\,a \sqrt{\frac{2}{\lambda}}\,\frac{1}{\sinh\left(\frac{\pi M}{\sqrt{2} m} \theta\right)}\,\,\,.
\EN
By making now the analitical continuation $\theta \rightarrow i \pi - \theta$ and using the above definitions 
(\ref{definitiong}), we arrive to   
\EQ
F_{-a,a}(\theta) \,= \,
\langle a\,\mid \varphi(0)\,\mid K_{-a,a}(\theta_1) K_{a,-a}(\theta_2) \rangle 
\,\propto \,\,  
\frac{1}{\sinh\left(\frac{(i \pi - \theta)}{\xi}\right)} \,\,\,.
\label{FFphi4}
\EN 
The poles of the above expression are located at 
\begin{equation}
\theta_{n}\,=\,i \pi \left(1 - \xi \,n\right)
\,\,\,\,\,\,\,
,
\,\,\,\,\,\,\,
n = 0,\pm 1,\pm 2,\ldots
\label{polesphi4}
\end{equation}
and, if 
\EQ
\xi \geq 1 \,\,\,,
\label{conditionphi4}
\EN 
none of them is in the physical strip $0 < {\rm Im}\,\theta < \pi$. Consequently, in the 
range of the coupling constant  
\EQ
\frac{\lambda}{m^2} \geq \frac{2 \pi}{3} \,\frac{1}{1 + \pi c} \,=\,1.02338...
\label{criticalg}
\EN 
the theory does not have any neutral bound states, neither on the vacuum to the right nor on the one to 
the left. Viceversa, if $\xi < 1$, there are $n = \left[\frac{1}{\xi}\right]$ neutral bound states, 
where $[ x ]$ denote the integer part of the number $x$. Their semiclassical masses are 
given by 
\begin{equation}
m_{b}^{(n)} \,=\, 2 M\,\sin\left[n\frac{\pi \xi}{2}\,\right]\,=\,
n\,\,m_b\left[1 - \frac{3}{32}\,\frac{\lambda^2}{m^4} 
\,n^{2} +...\right]\,.
\label{massphi4}
\end{equation}
Note that the leading term is given by multiples of the mass of the elementary boson $\mid B_1\rangle$.  
Therefore the $n$-th breather may be considered as a loosely bound state of $n$ of it, with the binding 
energy provided by the remaining terms of the above expansion. But, for the non-integrability of the 
theory, all particles with mass $m_n > 2 m_1$ will eventually decay. It is easy to see that, if there 
are at most two particles in the spectrum, it is always valid the inequality $m_2 < 2 m_1$. However, 
if $\xi < \frac{1}{3}$, for the higher particles one always has 
\EQ
m_k > 2 m_1
\,\,\,\,\,\,\,
,
\,\,\,\,\,\,\,
{\makebox for}\,\, k=3,4,\ldots n \,\,\,.
\EN 
According to the semiclassical analysis, the spectrum of neutral particles of $\varphi^4$ theory is then 
as follows: (i) if $\xi > 1$, there are no neutral particles; (ii) if $\frac{1}{2} < \xi < 1$, there is one 
particle; (iii) if $\frac{1}{3} < \xi < \frac{1}{2}$ there are two particles; (iv) if $\xi < \frac{1}{3}$ 
there are $\left[\frac{1}{\xi}\right]$ particles, although only the first two are stable, because the others 
are resonances.  
\vspace{5mm}

\begin{figure}[h]
\hspace{45mm}
\vspace{3mm}
\psfig{figure=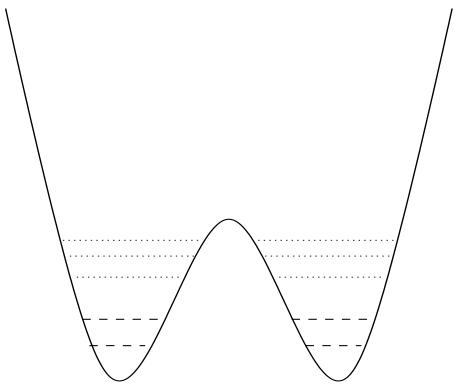,height=5cm,width=6cm}
\vspace{1mm}
\caption{{\em Neutral bound states of $\varphi^4$ theory for $g < 1$. The lowest two lines are 
the stable particles whereas the higher lines are the resonances.}}
\label{doublewell}
\end{figure}

Let us now briefly mention some general features of the semiclassical methods, starting from 
an equivalent way to derive the Fourier transform of the kink solution. To simplify the notation, 
let's get rid of all possible constants and consider the Fourier transform of the derivative of 
the kink solution, expressed as 
\EQ
G(k) \,=\,\int_{-\infty}^{\infty} d x \,e^{i k x} \frac{1}{\cosh^2 x} \,\,\,. 
\EN 
We split the integral in two terms
\EQ
G(k) \,=\,\int_{-\infty}^0 dx \,e^{i k x} \,\frac{1}{\cosh^2 x} 
+ \int_0^{\infty} dx \,e^{i k x} \,\frac{1}{\cosh^2 x} 
\,\,\,,
\label{FT}
\EN 
and we use the following series expansion of the integrand, valid on the entire real axis (except the origin)  
\EQ
\frac{1}{\cosh^2 x} \,=\,4 \,\sum_{n=1}^{\infty} (-1)^{n+1} n \,e^{-2 n |x|} \,\,\,.
\EN 
Substituting this expression into (\ref{FT}) and computing each integral, we have 
\EQ
G(k) \,=\,4 i\sum_{n=1}^{\infty} (-1)^{n+1} n \left[-\frac{1}{i k + 2n} + \frac{1}{-i k + 2n }\right] 
\,\,\,. 
\EN 
Obviously it coincides with the exact result, $G(k) \,=\,\pi k/\sinh\frac{\pi}{2} k$, but this derivation 
permits to easily interpret the physical origin of each pole. In fact, changing $k$ to the original variable
in the crossed channel, $k \rightarrow (i \pi - \theta)/\xi$, we see that the poles which determine the bound 
states at the vacuum $\mid a \rangle$ are only those relative to the exponential behaviour of the kink 
solution at $x \rightarrow -\infty$. This is precisely the point where the classical kink solution takes 
values on the vacuum $\mid a \rangle$. In the case of $\varphi^4$, the kink and the antikink are the same 
function (up to a minus sign) and therefore they have the same exponential approach at $x = -\infty$ at both 
vacua $\mid \pm 1 \rangle$. Mathematically speaking, this is the reason for the coincidence of the 
bound state spectrum on each of them: this does not necessarily happens in other cases, as 
we will see in the next section, for instance. 

The second comment concerns the behavior of the kink solution near the minima of the potential. 
In the case of $\varphi^4$, expressing the kink solution as 
\EQ
\varphi(x) \,=\,\frac{m}{\sqrt \lambda} \,\tanh\left[\frac{m \,x}{\sqrt 2}\right] \,=\,
\frac{m}{\sqrt \lambda}\,\,\frac{e^{\sqrt{2} \,x} -1}{e^{\sqrt{2}\,x} + 1} \,\,\,,
\EN 
and expanding around $x = -\infty$, we have 
\EQ
\varphi(t) \,=\,-\frac{m}{\sqrt \lambda}\,\left[1 - 2 t + 2 t^2 - 2 t^3 + \cdots 2 \,(-1)^n t^n \cdots \right] 
\,\,\,,
\EN 
where $t = \exp[\sqrt{2} x]$. Hence, all the sub-leading terms are exponential factors, with exponents which 
are multiple of the first one. Is this a general feature of the kink solutions of any theory? It can be 
proved that the answer is indeed positive \cite{kink}.

The fact that the approach to the minimum of the kink solutions is always through multiples of the 
same exponential (when the curvature $\omega$ at the minimum is different from zero) implies 
that the Fourier transform of the kink solution has poles regularly spaced by $\xi_a \equiv 
\frac{\omega}{\pi M_{ab}}$ in the variable $\theta$. If the first of them is within the physical 
strip, the semiclassical mass spectrum derived from the formula (\ref{remarkable1}) near the vacuum 
$\mid a \,\rangle$ has therefore the universal form 
\EQ
m_n \,=\,2 M_{ab} \,\sin\left(n\,\frac{\pi\,\xi_a}{2}\right) \,\,\,. 
\EN 
As we have previously discussed, this means that, according to the value of $\xi_a$, 
we can have only the following situations at the vacuum $\mid a \,\rangle$: (a) no bound state if
$ \xi_a > 1$; (b)  one particle if $\frac{1}{2} < \xi_a < 1$; (c) two particles 
if $\frac{1}{3} < \xi_a < \frac{1}{2}$; (d) $\left[\frac{1}{\xi_a}\right]$ particles 
if $\xi_a < \frac{1}{3}$, although only the first two are stable, the others being 
resonances. So, semiclassically, each vacuum of the theory cannot have more than two stable particles above it. 
Viceversa, if $\omega = 0$, there are no poles in the Fourier transform of the kink and 
therefore there are no neutral particles near the vacuum $\mid a \,\rangle$.

\section{Asymmetric wells}\label{phi6section}

In order to have a polynomial potential with two asymmetric wells, one must necessarily employ 
higher powers than $\varphi^4$. The simplest example of such a potential is obtained with a 
polynomial of maximum power $\varphi^6$, and this is the example discussed here. Apart from its simplicity, 
the $\varphi^6$ theory is relevant for the class of universality of the Tricritical Ising Model \cite{multi}. As 
we can see, the information available on this model will turn out to be a nice confirmation of the semiclassical 
scenario. . 

\begin{figure}[h]
\hspace{45mm}
\vspace{10mm}
\psfig{figure=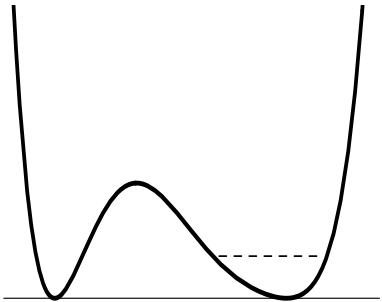,height=4cm,width=5cm}
\vspace{1mm}
\caption{{\em Example of $\varphi^6$ potential with two asymmetric wells and a bound state only on one of them.}}
\label{potential6}
\end{figure}

A class of potentials which may present two asymmetric wells is given by 
\EQ
U(\varphi) \,=\,\frac{\lambda}{2}\,\left(\varphi + a\frac{m}{\sqrt{\lambda}}\right)^2 \,
\left(\varphi - b\frac{m}{\sqrt{\lambda}}\right)^2 \,\left(\varphi^2 + c \frac{m^2}{\lambda}\right) \,\,\,, 
\label{phi6}
\EN 
with $a,b,c$ all positive numbers. To simplify the notation, it is convenient to use the 
dimensionless quantities obtained by rescaling the coordinate as $x^{\mu} \rightarrow  m x^{\mu}$ 
and the field as $\varphi(x) \rightarrow \sqrt{\lambda}/m \varphi(x)$. In this way the 
lagrangian of the model becomes 
\EQ
{\cal L} \,=\,\frac{m^6}{\lambda^2} \left[\frac{1}{2} (\partial \varphi)^2 - 
\frac{1}{2} (\varphi+a)^2 (\varphi-b)^2 
(\varphi^2 + c) \right]\,\,\,.
\label{newphi6}
\EN 
The minima of this potential are localised at $\varphi_0^{(0)} = - a$ and $\varphi_1^{(0)} = b$ and 
the corresponding ground states will be denoted by $\mid 0 \,\rangle$ and $\mid 1 \,\rangle$. The 
curvature of the potential at these points is given by 
\EQ
\begin{array}{lll}
U''(-a) & \equiv & \omega^2_0 = (a+b)^2 (a^2 + c) \,\,\,;\\
U''(b) & \equiv & \omega^2_1 = (a+b)^2 (b^2 + c)\,\,\,.
\end{array}
\label{curvature}
\EN 
For $ a \neq b$, we have two asymmetric wells, as shown in Figure \ref{potential6}. To be 
definite, let's assume that the curvature at the vacuum $\mid 0\,\rangle$ is higher than the 
one at the vacuum $\mid 1\,\rangle$, i.e. $a > b$.

\vspace{3mm}

The problem we would like to examine is whether the spectrum of the neutral particles $\mid B \,\rangle_{s}$ 
($ s = 0,1$) may be different at the two vacua, in particular, whether it would be possible that one of them 
(say $\mid 0 \rangle$) has no neutral excitations, whereas the other has just one neutral 
particle. The ordinary perturbation theory shows that both vacua has neutral excitations, although 
with different value of their mass: 
\EQ
m^{(0)} \,= \,(a+b) \sqrt{2 \, (a^2 + c)} 
\,\,\,\,\,\,\,
,
\,\,\,\,\,\,\,
m^{(1)} \,=\, (a+b) \sqrt{2 \, (b^2 + c)} 
\,\,\,.
\label{baremasses}
\EN 

\begin{figure}[h]
\hspace{30mm}
\vspace{1mm}
\psfig{figure=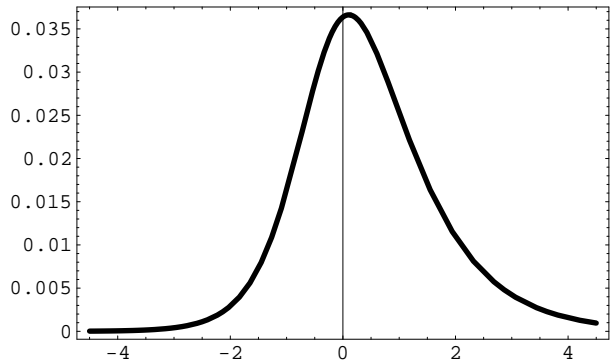,height=5cm,width=8cm}
\vspace{1mm}
\caption{{\em Typical shape of $\left(\frac{d \varphi}{dx}\right)_{01}$, obtained by a numerical solution of eq.\,
(\ref{kinkphi6}).}}
\label{asymmsol}
\end{figure}

Let's see, instead, what is the semiclassical scenario. The kink equation is given in this case by 
\EQ
\frac{d\varphi}{d x} \,=\,\pm (\varphi + a) (\varphi - b) \,\sqrt{\varphi^2 + c}\,\,\,.
\label{kinkphi6}
\EN 
We will not attempt to solve exactly this equation but we can present nevertheless its main features. 
The kink solution interpolates between the values $-a$ (at $ x = -\infty$) and $b$ (at $x = +\infty$). The 
anti-kink solution does viceversa, but with an important difference: its behaviour at $x = -\infty$ 
is different from the one of the kink. As a matter of fact, the behaviour at $x = - \infty$ of the kink 
is always equal to the behaviour at $x = +\infty$ of the anti-kink (and viceversa), but the two vacua are 
approached, in this theory, differently. This is explicitly shown in Figure \ref{asymmsol} and proved  
in the following.


Let us consider the limit $x \rightarrow - \infty$ of the kink solution. For these large values of $x$, 
we can approximate eq.\,(\ref{kinkphi6}) by substituting, in the second and in the 
third term of the right-hand side, $\varphi \simeq -a$, with the result 
\EQ
\left(\frac{d\varphi}{dx}\right)_{0,1} \simeq (\varphi + a) (a + b) \sqrt{a^2 + c} 
\,\,\,\,\,
,
\,\,\,\,\,
x \rightarrow - \infty
\EN 
This gives rise to the following exponential approach to the vacuum $\mid 0 \rangle$ 
\EQ
\varphi_{0,1}(x) \simeq -a + A \exp(\omega_0 x) 
\,\,\,\,\,\,\,
,
\,\,\,\,\,\,\,
x \rightarrow - \infty 
\EN 
where $A > 0$ is a arbitrary costant (its actual value can be fixed by properly solving the non-linear 
differential equation). To extract the behavior at $x \rightarrow -\infty$ of the anti-kink, we 
substitute this time $\varphi \simeq b$ into the first and third term of the right hand side of 
(\ref{kinkphi6}), so that 
\EQ
\left(\frac{d\varphi}{d x}\right)_{1,0} \simeq (\varphi - b) (a + b) \sqrt{b^2 + c} 
\,\,\,\,\,\,
,
\,\,\,\,\,\,
x \rightarrow - \infty
\EN 
This ends up in the following exponential approach to the vacuum $\mid 1 \rangle$ 
\EQ
\varphi_{1,0}(x) \simeq b - B \exp(\omega_1 x) 
\,\,\,\,\,\,\,\,
,
\,\,\,\,\,\,\,\,
x \rightarrow - \infty
\EN 
where $B > 0$ is another constant. Since $\omega_0 \neq \omega_1$, the asymptotic behaviour of the two solutions 
gives rise to the following poles in their Fourier transform 
\begin{eqnarray}
{\cal F}(\varphi_{0,1}) & \rightarrow & \frac{A}{\omega_0 + i k} \nonumber \\
& & \label{polephi6}\\
{\cal F}(\varphi_{1,0}) & \rightarrow & \frac{-B}{\omega_1 + i k} \nonumber 
\end{eqnarray}
In order to locate the pole in $\theta$, we shall reintroduce the correct units. Assuming to have solved  
the differential equation (\ref{kinkphi6}), the integral of its energy density gives the common mass of the kink 
and the anti-kink. In terms of the constants in front of the Lagrangian (\ref{newphi6}), its value is given by 
\EQ
M\,=\,\frac{m^5}{\lambda^2} \,\alpha \,\,\,,
\EN 
where $\alpha$ is a number (typically of order $1$), coming from the integral of the adimensional energy 
density (\ref{integralmass}). Hence, the first pole\footnote{In order to determine the others, one should 
look for the subleading exponential terms of the solutions.} of the Fourier transform of the kink and the 
antikink solution are localised at 
\begin{eqnarray}
\theta^{(0)} \,& \simeq & \,i\pi \left( 1 - \omega_0 \,\frac{m}{\pi M}\right) = i\pi 
\left(1 - \omega_0 \,\frac{\lambda^2}{\alpha m^4}\right)
\nonumber \\
& & \\
\theta^{(1)} \,& \simeq & \,i\pi \left( 1- \omega_1 \,\frac{m}{\pi M}\right) 
\,=\, i\pi \left(1 - \omega_1 \,\frac{\lambda^2}{\alpha m^4}\right) \nonumber 
\end{eqnarray}
If we now choose the coupling constant in the range 
\EQ
\frac{1}{\omega_0} < \frac{\lambda^2}{m^4} < \frac{1}{\omega_1}    \,\,\,,
\label{range}
\EN 
the first pole will be out of the physical sheet whereas the second will still remain inside it!  
Hence, the theory will have only one neutral bound state, localised at the vacuum $\mid 1 \,\rangle$. 
This result may be expressed by saying that the appearance of a bound state depends on the 
order in which the topological excitations are arranged: an antikink-kink configuration gives rise 
to a bound state whereas a kink-antikink does not. 

Finally, notice that the value of the adimensional coupling constant can be chosen so that the mass 
of the bound state around the vacuum $\mid 1 \,\rangle$ becomes equal to mass of the kink. 
This happens when 
\EQ
\frac{\lambda^2}{m^4} \,=\,\frac{\alpha}{3 \omega_1} \,\,\,.
\EN  

Strange as it may appear, the semiclassical scenario is well confirmed by an explicit example. 
This is provided by the exact scattering theory of the Tricritical Ising Model perturbed by its sub-leading 
magnetization. Firstly discovered through a numerical analysis of the spectrum of this model \cite{LMC}, its 
exact scattering theory has been discussed later in \cite{CKM}.

\section{Conclusions}

In this paper we have used simple arguments of the semi-classical analysis to investigate the 
spectrum of neutral particles in quantum field theories with kink excitations. We have 
concentrated our analysis on two cases: the first relative to a potential with symmetric wells, 
the second concerning with a potential with asymmetric wells. Leaving apart the 
exact values of the quantities extracted by the semiclassical methods, it is perhaps more important 
to underline some general features which have emerged through this analysis. One of them concerns, 
for instance, the existence of a critical value of the coupling constant, beyond which there are 
no neutral bound states. Another result is about the maximum number $n \leq 2$ of neutral particles 
living on a generica vacuum of a non-integrable theory. An additional aspect is the role played by 
the asymmetric vacua, which may have a different number of neutral excitations
above them.

\vspace{1cm}
\begin{flushleft}\large
\textbf{Acknowledgements}
\end{flushleft}

I would like to thank G. Delfino and V. Riva for interesting discussions. 
I am particularly grateful to M. Peyrard for very useful and enjoyable 
discussions on solitons. This work was done under partial support of 
the ESF grant INSTANS.



\begin{thebibliography}{1}
\bibitem{zamzam} A.B. Zamolodchikov and Al.B. Zamolodchikov,
Ann. Phys. 120 (1979) 253.
\bibitem{Zam} A.B. Zamolodchikov, 
Adv. Stud. Pure Math. 19 (1989), 641. 
\bibitem{Smirnov} F. A. Smirnov, {\em Form Factors in Completely 
Integrable Models of Quantum Field Theory}, (World Scientific, 
Singapore, 1992); M. Karowski and P. Weisz, {\em Nucl. Phys.} {\bf D 139}, (1978), 455.
\bibitem{DMIsing} G. Delfino and G. Mussardo, {\em Nucl. Phys.} {\bf B 455}, (1995), 724;
G. Delfino and P. Simonetti, {\em Phys. Lett.} {\bf B 383}, (1996), 450.
\bibitem{GM} G. Mussardo,
{\em Phys. Rept.} {\bf 218} (1992), 215.
\bibitem{zam.massless} Al.B.Zamolodchikov, {\em Nucl.Phys.} {\bf B 358}, (1991), 524.
\bibitem{zam-zam.massless} A.B.Zamolodchikov and Al.B.Zamolodchikov,   
{\em Nucl.Phys.} {\bf B 379}, (1992), 602.
\bibitem{fsw}P. Fendley, H. Saleur and N.P. Werner, {\em Nucl.Phys.} {\bf B 430}, (1994), 577.  
\bibitem{dms2} G.Delfino, G.Mussardo and P.Simonetti, {\em Phys. Rev.} {\bf D 51}, (1995), 6622. 
\bibitem{dms} G.Delfino, G.Mussardo and P.Simonetti,{\em Nucl.Phys.} {\bf B 473}, (1996), 469.  
\bibitem{grinza} P. Grinza, G. Delfino and G. Mussardo, hep/th 0507133, {\em Nucl. Phys. B} in press.
\bibitem{Aldo} G. Delfino, {\em Particle decay in Ising field theory with magnetic field}, 
Proceedings ICMP 2006. 
\bibitem{mccoy} B.M. McCoy and T.T. Wu, {\em Phys. Rev.} {\bf D 18} (1978), 1259.
\bibitem{fonsecazam}P. Fonseca and A.B. Zamolodchikov, {\em J.Stat.Phys.}{\bf 110} (2003), 527.
\bibitem{rut}S.B. Rutkevich, {\em Phys. Rev. Lett.} {\bf 95} (2005), 250601.
\bibitem{zamfon} P. Fonseca and A.B. Zamolodchikov, {\em Ising Spectoscopy I: 
Mesons at $T < T_c$}, hep-th/0612304.
\bibitem{dm} G. Delfino and G. Mussardo, {\em Nucl. Phys.} {\bf B 516}, (1998), 675.
\bibitem{takacs} Z. Bajnok, L. Palla, G. Takacs, F. Wagner, {\em Nucl.Phys.} {\bf B 601}, (2001), 503.
\bibitem{CMPRL} D. Controzzi and G. Mussardo, {\em Phys. Rev. Lett.} {\bf 92}, (2004), 021601.
\bibitem{conmus} D. Controzzi and G. Mussardo, {\em Phys. Lett.} {\bf B 617}, (2005), 133. 
\bibitem{decay} G. Delfino, P. Grinza and G. Mussardo,  {\em Nucl. Phys.} {\bf B 737} (2006), 291. 

\bibitem{DHN} R.F.Dashen, B.Hasslacher and A.Neveu,  {\em Phys. Rev.} {\bf D 10}
(1974) 4130;

R.F.Dashen, B.Hasslacher and A.Neveu, {\em Phys. Rev.} {\bf D 11} (1975) 3424.

\bibitem{GJ} J. Goldstone and R. Jackiw, {\em Phys.Rev.} {\bf D 11} (1975) 1486.

\bibitem{JW} R. Jackiw and G. Woo, {\em Phys. Rev.} {\bf D 12} (1975), 1643. 

\bibitem{FFvolume} G. Mussardo, V. Riva and G. Sotkov,  
{\em Nucl. Phys.} {\bf B 670} (2003), 464. 

\bibitem{volume} G. Mussardo, V. Riva and G. Sotkov, 
{\em Nucl. Phys.} {\bf B 699} (2004), 545. 

G. Mussardo, V. Riva and G. Sotkov, 
{\em Nucl. Phys.} {\bf B 705} (2005), 548

\bibitem{kink} G. Mussardo, {\em Neutral bound states in kink-like theories}, 
hep-th/0607025, to appear on Nucl. Phys. B. 

\bibitem{multi} A.B. Zamolodchikov, 
{\em Sov.J.Nucl.Phys.} {\bf 44} (1986), 529.
 
\bibitem{LMC} M. Lassig, G. Mussardo and J.L. Cardy,  
{\em Nucl. Phys.} {\bf B 348} (1991), 591. 

\bibitem{CKM} F. Colomo, A. Koubek and G. Mussardo, 
{\em Int. Journ. Mod. Phys.} {\bf A 7} (1992), 5281.




\end{thebibliography}
\end{document}